%% file: 0main.tex
\documentclass[longtitle,preprint,5p,times,twocolumn]{elsarticle}

\usepackage{amssymb}

\usepackage{caption}
\usepackage{xspace}
\usepackage{longtable}
\usepackage{pdflscape}
\usepackage{multirow}
\usepackage{ragged2e}
\usepackage{booktabs}
\usepackage{hyperref}
\usepackage{lineno}
\usepackage[usenames,dvipsnames]{color}
\def\notenumber{ecce-paper-phys-2022-09}
\def\noteversion{v1.0}

\begin{document}
\begin{frontmatter}

\title{ECCE unpolarized TMD measurements}
\tnotetext[t1]{
	    \includegraphics[scale=0.075]{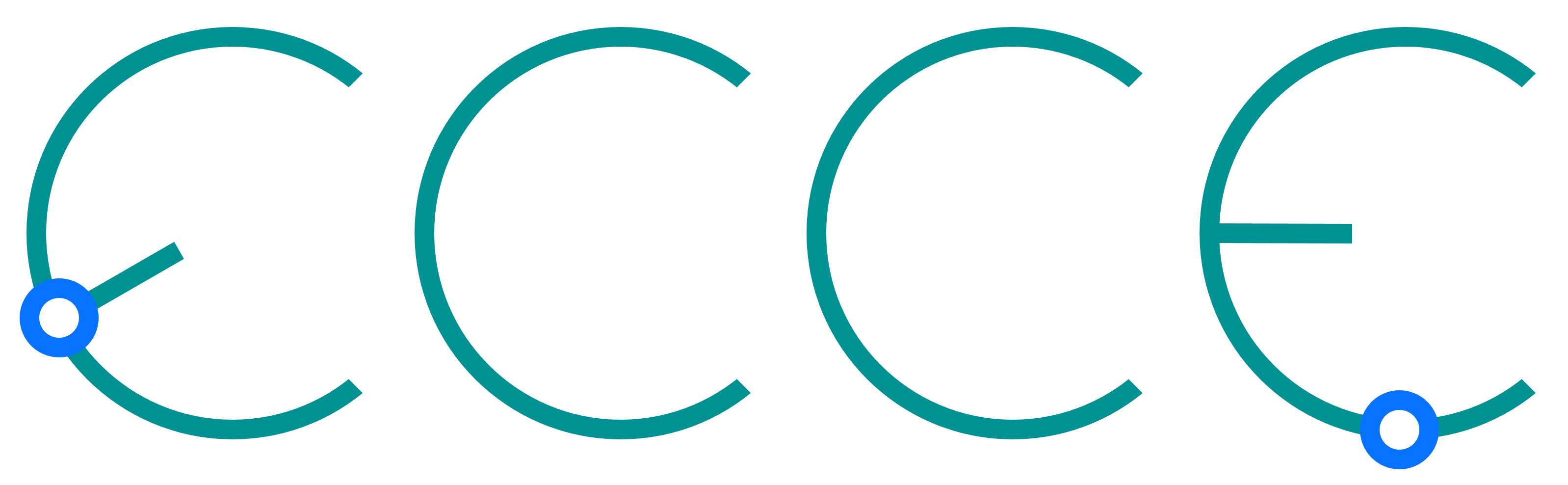} \\
	    \notenumber \\
	    \noteversion \\
	    \today	\\
}
\def\theaffn{\arabic{affn}} 
\input{extraauthors}
\input{authors-alphabetical}
\input{affiliations}
\input{extraaffiliations}

    \begin{abstract}
        \input{abstract}
    \end{abstract}
\date{\today}

\end{frontmatter}


\section {Introduction\label{sec:intro}}

\input{1intro}

\section {Simulations, data sets and selection criteria\label{sec:data}}
\input{2data}

\section{Transverse momentum dependent multiplicities}
\input{3mults}
\section{Transverse momentum dependent cross sections and projections}
\input{4xsecs}

\section{Impact studies}
\input{5impact}
\input{6outlook}
\bibliographystyle{unsrturl}
\bibliography{refs}

\end{document}

%% file: extraauthors.tex
\author[RIKEN,RBRC]{Ralf Seidl}
\author[Regensburg,Madrid]{Alexey Vladimirov}

%% file: authors-alphabetical.tex
%
%
%
%

\author[MoreheadState]{J.~K.~Adkins}
\author[RIKEN,RBRC]{Y.~Akiba}
\author[UKansas]{A.~Albataineh}
\author[ODU]{M.~Amaryan}
\author[Oslo]{I.~C.~Arsene}
\author[MSU]{C. Ayerbe Gayoso}
\author[Sungkyunkwan]{J.~Bae}
\author[UVA]{X.~Bai}
\author[BNL,JLab]{M.D.~Baker}
\author[York]{M.~Bashkanov}
% ...

%% file: affiliations.tex
%

\affiliation[AANL]{organization={A. Alikhanyan National Laboratory},
	 city={Yerevan},
	 country={Armenia}} 
 
\affiliation[AcademiaSinica]{organization={Institute of Physics, Academia Sinica},
	 city={Taipei},
	 country={Taiwan}} 
 
\affiliation[AUGIE]{organization={Augustana University},
	 city={Sioux Falls},
	 state={SD},
	 country={USA}} 
	 
\affiliation[BGU]{organizatoin={Ben-Gurion University of the Negev}, 
      city={Beer-Sheva},
      country={Israel}}

\affiliation[BNL]{organization={Brookhaven National Laboratory},
	 city={Upton},
	 state={NY},
	 country={USA}} 
 
\affiliation[BrunelUniversity]{organization={Brunel University London},
	 city={Uxbridge},
	 country={UK}} 
 
\affiliation[CanisiusCollege]{organization={Canisius College},
	 city={Buffalo},
	 state={NY},
	 country={USA}} 
 
\affiliation[CCNU]{organization={Central China Normal University},
	 city={Wuhan},
	 country={China}} 
 
\affiliation[Charles]{organization={Charles University},
	 city={Prague},
	 country={Czech Republic}} 
 
\affiliation[CIAE]{organization={China Institute of Atomic Energy, Fangshan},
	 city={Beijing},
	 country={China}} 
 
\affiliation[CNU]{organization={Christopher Newport University},
	 city={Newport News},
	 state={VA},
	 country={USA}} 
 
\affiliation[Columbia]{organization={Columbia University},
	 city={New York},
	 state={NY},
	 country={USA}} 
 
\affiliation[CUA]{organization={Catholic University of America},
	 city={Washington DC},
	 country={USA}} 
 
\affiliation[CzechTechUniv]{organization={Czech Technical University},
	 city={Prague},
	 country={Czech Republic}} 
 
\affiliation[Duquesne]{organization={Duquesne University},
	 city={Pittsburgh},
	 state={PA},
	 country={USA}} 
 
\affiliation[Duke]{organization={Duke University},
	 cite={Durham},
	 state={NC},
	 country={USA}} 
 
\affiliation[FIU]{organization={Florida International University},
	 city={Miami},
	 state={FL},
	 country={USA}} 
 
\affiliation[GeorgiaState]{organization={Georgia State University},
	 city={Atlanta},
	 state={GA},
	 country={USA}} 
 
\affiliation[Glasgow]{organization={University of Glasgow},
	 city={Glasgow},
	 country={UK}} 
 
\affiliation[GSI]{organization={GSI Helmholtzzentrum fuer Schwerionenforschung GmbH},
	 city={Darmstadt},
	 country={Germany}} 
 
\affiliation[GWU]{organization={The George Washington University},
	 city={Washington, DC},
	 country={USA}} 
 
\affiliation[Hampton]{organization={Hampton University},
	 city={Hampton},
	 state={VA},
	 country={USA}} 
 
\affiliation[HUJI]{organization={Hebrew University},
	 city={Jerusalem},
	 country={Isreal}} 
 
\affiliation[IJCLabOrsay]{organization={Universite Paris-Saclay, CNRS/IN2P3, IJCLab},
	 city={Orsay},
	 country={France}} 
	 
\affiliation[CEA]{organization={IRFU, CEA, Universite Paris-Saclay},
     cite= {Gif-sur-Yvette},
     country={France}
}

\affiliation[IMP]{organization={Chinese Academy of Sciences},
	 city={Lanzhou},
	 country={China}} 
 
\affiliation[IowaState]{organization={Iowa State University},
	 city={Iowa City},
	 state={IA},
	 country={USA}} 
 
\affiliation[JazanUniversity]{organization={Jazan University},
	 city={Jazan},
	 country={Sadui Arabia}} 
 
\affiliation[JLab]{organization={Thomas Jefferson National Accelerator Facility},
	 city={Newport News},
	 state={VA},
	 country={USA}} 
 
\affiliation[JMU]{organization={James Madison University},
	 city={Harrisonburg},
	 state={VA},
	 country={USA}} 
 
\affiliation[KobeUniversity]{organization={Kobe University},
	 city={Kobe},
	 country={Japan}} 
 
\affiliation[Kyungpook]{organization={Kyungpook National University},
	 city={Daegu},
	 country={Republic of Korea}} 
 
\affiliation[LANL]{organization={Los Alamos National Laboratory},
	 city={Los Alamos},
	 state={NM},
	 country={USA}} 
 
\affiliation[LBNL]{organization={Lawrence Berkeley National Lab},
	 city={Berkeley},
	 state={CA},
	 country={USA}} 
 
\affiliation[LehighUniversity]{organization={Lehigh University},
	 city={Bethlehem},
	 state={PA},
	 country={USA}} 
 
\affiliation[LLNL]{organization={Lawrence Livermore National Laboratory},
	 city={Livermore},
	 state={CA},
	 country={USA}} 
 
\affiliation[MoreheadState]{organization={Morehead State University},
	 city={Morehead},
	 state={KY},
	 }
 
\affiliation[MIT]{organization={Massachusetts Institute of Technology},
	 city={Cambridge},
	 state={MA},
	 country={USA}} 
 
\affiliation[MSU]{organization={Mississippi State University},
	 city={Mississippi State},
	 state={MS},
	 country={USA}} 
 
\affiliation[NCKU]{organization={National Cheng Kung University},
	 city={Tainan},
	 country={Taiwan}} 
 
\affiliation[NCU]{organization={National Central University},
	 city={Chungli},
	 country={Taiwan}} 
 
\affiliation[Nihon]{organization={Nihon University},
	 city={Tokyo},
	 country={Japan}} 
 
\affiliation[NMSU]{organization={New Mexico State University},
	 city={Las Cruces},
	 state={NM},
	 country={USA}} 
 
\affiliation[NRNUMEPhI]{organization={National Research Nuclear University MEPhI},
	 city={Moscow},
	 country={Russian Federation}} 
 
\affiliation[NRCN]{organization={Nuclear Research Center - Negev},
	 city={Beer-Sheva},
	 country={Isreal}} 
 
\affiliation[NTHU]{organization={National Tsing Hua University},
	 city={Hsinchu},
	 country={Taiwan}} 
 
\affiliation[NTU]{organization={National Taiwan University},
	 city={Taipei},
	 country={Taiwan}} 
 
\affiliation[ODU]{organization={Old Dominion University},
	 city={Norfolk},
	 state={VA},
	 country={USA}} 
 
\affiliation[Ohio]{organization={Ohio University},
	 city={Athens},
	 state={OH},
	 country={USA}} 
 
\affiliation[ORNL]{organization={Oak Ridge National Laboratory},
	 city={Oak Ridge},
	 state={TN},
	 country={USA}} 
 
\affiliation[PNNL]{organization={Pacific Northwest National Laboratory},
	 city={Richland},
	 state={WA},
	 country={USA}} 
 
\affiliation[Pusan]{organization={Pusan National University},
	 city={Busan},
	 country={Republic of Korea}} 
 
\affiliation[Rice]{organization={Rice University},
	 city={Houston},
	 state={TX},
	 country={USA}} 
 
\affiliation[RIKEN]{organization={RIKEN Nishina Center},
	 city={Wako},
	 state={Saitama},
	 country={Japan}} 
 
\affiliation[Rutgers]{organization={The State University of New Jersey},
	 city={Piscataway},
	 state={NJ},
	 country={USA}}

\affiliation[CFNS]{organization={Center for Frontiers in Nuclear Science},
	 city={Stony Brook},
	 state={NY},
	 country={USA}} 
 
\affiliation[StonyBrook]{organization={Stony Brook University},
	 city={Stony Brook},
	 state={NY},
	 country={USA}} 
 
\affiliation[RBRC]{organization={RIKEN BNL Research Center},
	 city={Upton},
	 state={NY},
	 country={USA}} 
	 
\affiliation[SDU]{organizaton={Shandong University},
     city={Qingdao},
     state={Shandong},
     country={China}}
     
\affiliation[Seoul]{organization={Seoul National University},
	 city={Seoul},
	 country={Republic of Korea}} 
 
\affiliation[Sejong]{organization={Sejong University},
	 city={Seoul},
	 country={Republic of Korea}} 
 
\affiliation[Shinshu]{organization={Shinshu University},
         city={Matsumoto},
	 state={Nagano},
	 country={Japan}} 
 
\affiliation[Sungkyunkwan]{organization={Sungkyunkwan University},
	 city={Suwon},
	 country={Republic of Korea}} 
 
\affiliation[TAU]{organization={Tel Aviv University},
	 city={Tel Aviv},
	 country={Israel}} 

\affiliation[USTC]{organization={University of Science and Technology of China},
     city={Hefei},
     country={China}}

\affiliation[Tsinghua]{organization={Tsinghua University},
	 city={Beijing},
	 country={China}} 
 
\affiliation[Tsukuba]{organization={Tsukuba University of Technology},
	 city={Tsukuba},
	 state={Ibaraki},
	 country={Japan}} 
 
\affiliation[CUBoulder]{organization={University of Colorado Boulder},
	 city={Boulder},
	 state={CO},
	 country={USA}} 
 
\affiliation[UConn]{organization={University of Connecticut},
	 city={Storrs},
	 state={CT},
	 country={USA}} 
 
\affiliation[UNGeorgia]{organization={University of North Georgia},
     cite={Dahlonega}, 
     state={GA},
     country={USA}}
     
\affiliation[UH]{organization={University of Houston},
	 city={Houston},
	 state={TX},
	 country={USA}} 
 
\affiliation[UIUC]{organization={University of Illinois}, 
	 city={Urbana},
	 state={IL},
	 country={USA}} 
 
\affiliation[UKansas]{organization={Unviersity of Kansas},
	 city={Lawrence},
	 state={KS},
	 country={USA}} 
 
\affiliation[UKY]{organization={University of Kentucky},
	 city={Lexington},
	 state={KY},
	 country={USA}} 
 
\affiliation[Ljubljana]{organization={University of Ljubljana, Ljubljana, Slovenia},
	 city={Ljubljana},
	 country={Slovenia}} 
 
\affiliation[NewHampshire]{organization={University of New Hampshire},
	 city={Durham},
	 state={NH},
	 country={USA}} 
 
\affiliation[Oslo]{organization={University of Oslo},
	 city={Oslo},
	 country={Norway}} 
 
\affiliation[Regina]{organization={ University of Regina},
	 city={Regina},
	 state={SK},
	 country={Canada}} 
 
\affiliation[USeoul]{organization={University of Seoul},
	 city={Seoul},
	 country={Republic of Korea}} 
 
\affiliation[UTsukuba]{organization={University of Tsukuba},
	 city={Tsukuba},
	 country={Japan}} 
	 
\affiliation[UoT]{organization={University of Texas},
    city={Austin},
    state={Texas},
    country={USA}}
 
\affiliation[UTK]{organization={University of Tennessee},
	 city={Knoxville},
	 state={TN},
	 country={USA}} 
 
\affiliation[UVA]{organization={University of Virginia},
	 city={Charlottesville},
	 state={VA},
	 country={USA}} 
 
\affiliation[Vanderbilt]{organization={Vanderbilt University},
	 city={Nashville},
	 state={TN},
	 country={USA}} 
 
\affiliation[VirginiaTech]{organization={Virginia Tech},
	 city={Blacksburg},
	 state={VA},
	 country={USA}} 
 
\affiliation[VirginiaUnion]{organization={Virginia Union University},
	 city={Richmond},
	 state={VA},
	 country={USA}} 
 
\affiliation[WayneState]{organization={Wayne State University},
	 city={Detroit},
	 state={MI},
	 country={USA}} 
 
\affiliation[WI]{organization={Weizmann Institute of Science},
	 city={Rehovot},
	 country={Israel}} 
 
\affiliation[WandM]{organization={The College of William and Mary},
	 city={Williamsburg},
	 state={VA},
	 country={USA}} 
 
\affiliation[Yamagata]{organization={Yamagata University},
	 city={Yamagata},
	 country={Japan}} 
 
\affiliation[Yarmouk]{organization={Yarmouk University},
	 city={Irbid},
	 country={Jordan}} 
 
\affiliation[Yonsei]{organization={Yonsei University},
	 city={Seoul},
	 country={Republic of Korea}} 
 
\affiliation[York]{organization={University of York},
	 city={York},
	 country={UK}} 
 
\affiliation[Zagreb]{organization={University of Zagreb},
	 city={Zagreb},
	 country={Croatia}} 
 

%% file: extraaffiliations.tex

 
\affiliation[Regensburg]{organization={Universit"at Regensburg},
	 city={Regensburg},
	 	 postcode={},
	 country={Germany}} 

\affiliation[Madrid]{organization={Universidad Complutense de Madrid},
city={Madrid},
postcode={ E-28040},
country={Spain}}

%% file: abstract.tex
We performed feasibility studies for various measurements that are related to unpolarized TMD distribution and fragmentation functions. The processes studied include semi-inclusive Deep inelastic scattering (SIDIS) where single hadrons (pions and kaons) were detected in addition to the scattered DIS lepton. The single hadron cross sections and multiplicities were extracted as a function of the DIS variables $x$ and $Q^2$, as well as the semi-inclusive variables $z$, which corresponds to the momentum fraction the detected hadron carries relative to the struck parton and $P_T$, which corresponds to the transverse momentum of the detected hadron relative to the virtual photon. The expected statistical precision of such measurements is extrapolated to accumulated luminosities of 10 fb$^{-1}$ and potential systematic uncertainties are approximated given the deviations between true and reconstructed yields. 

%% file: 1intro.tex
The study of transverse momentum dependent distribution and fragmentation functions originated with the first nonzero single transverse spin asymmetries that were discovered in the past by the E704 experiment in fixed-target proton-proton collisions \cite{FNAL-E704:1991ovg}. Both of the most famous effects, the Sivers \cite{Sivers:1989cc} effect and the Collins \cite{Collins:1992kk} effect that were initially suggested to describe these asymmetries require an intrinsic transverse momentum dependence on the distribution and fragmentation sides. 
While those two effects are also explicitly spin dependent, unpolarized transverse momentum dependent, TMD, functions are furthermore of importance in many processes. From the low-$x$ processes where the transverse momentum dependence of the gluon distribution function may affect potential saturation effects to the transverse momentum dependence of PDFs that may affect the actual transverse momentum dependent cross sections of Higgs or heavy boson production at the LHC, TMDs play an important role.

For the most part the information on explicitly transverse momentum dependent distribution functions originates from Drell-Yan, DY, and heavy boson production measurements, predominantly at the Tevatron and the LHC, as well as fixed target DY. However, due to the nature of these processes, only a very limited knowledge on the flavor structure of TMDs can be obtained this way. On the other hand, in semi-inclusive deep inelastic scattering, SIDIS, measurements, one mostly obtains cross sections that are convolutions of TMD distribution and fragmentation functions. The fragmentation functions provide an additional flavor sensitivity that is neither available in the Drell-Yan type measurements nor in any jet type DIS measurements. 

At the moment the information on the transverse momentum dependent fragmentation functions is also rather limited, with essentially only one pure measurement from $e^+e^-$ annihilation \cite{Belle:2019ywy} available and a few SIDIS measurements from fixed target experiments \cite{HERMES:2012uyd,COMPASS:2017mvk}. 

Some permutations of all this data have been combined in global fits that try to extract the flavor and transverse momentum dependent distribution and fragmentation functions simultaneously. The most recent of those are refs.~ \cite{Scimemi:2019cmh,Bacchetta:2019sam,Boglione:2020auc,Kang:2020yqw}, but none have so far extracted the TMDs from all of this data simultaneously. 

Closely related are various theoretical questions that are not entirely answered, such as which regions in phase space can actually be interpreted via TMDs and how that may limit the available data \cite{Boglione:2019nwk}. 
The authors identify regions where TMD factorization should be applicable in their approach and regions where other treatments may be relevant, such as collinear factorization.
One finds, that at lower $x$ and scales $Q^2$, only very small transverse momenta can be interpreted that way while higher transverse momenta likely involve higher order collinear processes. At higher scales and $x$ more of the phase space appears to be applicable for TMD interpretations. The question which regions are applicable in various factorization frameworks is currently under active discussion among theorists, where the EIC data help cover a much larger phase space than so far. However, not only the TMD region is relevant, but also the collinear region and the transition between those are of great interest. There are calculations that show how those can be related \cite{Ji:2006ub,Boer:2011fx}. This relation is commonly used when taking into account TMD evolution of TMD related cross sections and asymmetries. However, some aspects of the scale evolution of TMDs are non-perturbative in itself and two fits mentioned above, despite having used the same data, come to rather different evolution effects of even the unpolarized TMD PDFs.


The EIC in general and ECCE in particular can help answer most of these questions as a very large range in phase space is covered from the low scales of most fixed target DIS experiments to the high scales of the DY and even close to some heavy boson production measurements. The particle identification capabilities of ECCE furthermore may allow the flavor decomposition of TMDs and may answer the question whether valence and sea quark intrinsic transverse momenta in the nucleon are distributed differently. Also the regions of applicability of TMDs and collinear PDFs/FFs can be explored in detail. 

Naturally, any unpolarized TMDs are also the baseline for any polarized TMDs such the as the Sivers and Collins functions and the related Tensor charge of the nucleon. So, improving the knowledge and precision on the unpolarized TMDs in turn will improve the precision of these polarized functions as well.

%% file: 2data.tex
\subsection{Data sample}
The simulated data was obtained using the pythiaeRHIC implementation of {\sc pythia}6 with the same settings and events that were also used in the SIDIS studies of the EIC Yellow report \cite{EICYellowReport} \footnote{The generated data, as well as the steering files are available under {\it /gpfs02/eic/DATA/YR\_SIDIS/} at RCF as well as shared via the BNL Box service.}. The generated data, in its eic-smear file format was then run through a {\sc geant}4 simulation of the ECCE detector that contains all the relevant tracking detectors and calorimeters, as well as some of the support material, magnet yoke, the PID detectors, etc. The truth and reconstructed data were then analyzed to obtain the unpolarized TMD cross sections. In the reconstructed data the {\sc geant} output included the simulated detector response but for the most part not yet actual digitization in readout electronics. More details on these simulations can be found in \cite{ecce-sim}.
The PID information in these simulations came from a parametrization based on the rapidity and momentum dependent PID resolutions that can be expected for the various PID subsystems.   

The data was obtained at the energy combinations that are summarized in Table \ref{tab:lumi} where the simulations for low and high $Q^2$ were created separately in order to obtain sufficient statistics at higher $Q^2$. 
All collision energies are analyzed separately and the pseudo-data is stored separately for each particle type and energy combination. Unlike in the Yellow report, no dedicated e+$^3$He
 simulations were run and instead the yellow report uncertainties were re-scaled based on the ECCE e+p simulations. 
 As can be seen from the luminosities in the table, especially at low $Q^2$ the accumulated data is still far below the level of statistics to be expected from the EIC. Nevertheless the statistics are large enough to evaluate the statistical uncertainties that can be expected except at the borders of phase space (particularly at high-$z$ and high $P_T$). At the higher $Q^2>100$ GeV$^2$ range the lumionsities are generally larger which in turn compensates for the lower cross sections and event rates expected there. 
 
\begin{table}[htb]
    \centering
    \begin{tabular}{c | c  | r | c }
        Energy &  $Q^2$ range & events & Luminosity (fb$^{-1}$) \\\hline
          18x275 & 1 - 100 & 38.71M & 0.044 \\
               &  $>$  100 & 3.81M & 1.232 \\   
        18x100 & 1 - 100 & 14.92M & 0.022 \\
               &  $>$  100 & 3.72M & 2.147 \\
        10x100 & 1 - 100 & 39.02M & 0.067 \\
               &  $>$  100 & 1.89M & 1.631 \\
         5x41  & 1 - 100 & 39.18M & 0.123 \\
               &  $>$  100 & 0.96M & 5.944 \\\hline
    \end{tabular}
    \caption{MC statistics and luminosities used for the Single spin asymmetry simulations. Part of the lower $Q^2$ range data was obtained from simulations without upper $Q^2$ cut.}
    \label{tab:lumi}
\end{table}

\subsection{Event and hadron selection}

For the true as well as reconstructed events typical DIS selection criteria were applied as the following:
\begin{eqnarray}
Q^2 & > & 1 {\mathrm GeV}^2 \\ 
0.01  & < y < & 0.95 \\
W^2 & > & 10 {\mathrm GeV}^2 \quad ,
\end{eqnarray}
where the momentum transfer $Q^2$ needs to be large enough to allow a perturbative QCD description of the hard process. The invariant mass of the hadronic final state $W^2$ removes contributions from nucleon resonances from the measurements. The selection criteria on the inelasticity $y$ limit the ranges of large radiative contributions as well as regions where the reconstruction of the DIS kinematics via the scattered lepton creates large uncertainties. 
Particularly at low $Q^2$ and $W^2$ values the smearing that happens when using the reconstructed variables will move events outside of the accepted kinematic range and thus introduced some inefficiency. At higher scales that effect is less pronounced although the smearing can be still sizeable. At present this analysis reports only the results using the scattered lepton to determine the DIS and SIDIS kinematics. Both true and reconstructed kinematics are considered and will be compared to obtain a rough measure of systematic uncertainties.

For the SIDIS events, no explicit selection criteria in $z$ or $P_T$ were introduced, which means that particularly at low $z$ contributions from target fragmentation may be present as well. 
At present charged pions, kaons and protons are analyzed where the true particle information has been used, assuming that the anyway only moderate particle mis-identification will be unfolded in the actual ECCE data. 

\subsubsection{Binning}
Similar to many SIDIS related studies in the Yellow report two types of binnings have been used in these studies. For the unpolarized TMD measurements where no additional binning in azimuthal angles is required a slightly finer binning in $x$ and $Q^2$ was selected. In $x$ 5 logarithmically equidistant bins per decade were selected, namely [0.1, 0.158489, 0.251189, 0.398107, 0.630957, 1.0] and similarly for the decades down to $10^{-5}$, resulting in 25 bins in total. For $Q^2$, the binning consists of 4 bins per decade, namely [1, 1.77828, 3.16228, 5.62341, 10.] and similarly up to 1000 GeV$^2$. Above 1000 GeV$^2$ only one bin was assigned as only the highest collision energies can reach it and cross sections are very low already. 
 The bin boundaries are also summarized in Table \ref{tab:binning}. Additionally, in each kinematic $x,Q^2$ bin, the events are put into bins of the fractional energy $z$ and the transverse momentum of the detected hadron relative to the virtual photon direction in the proton center-of-mass system.

\begin{table*}[h]
    \centering
    \begin{tabular}{c|c}
    Kinematic variable & Bin boundaries \\ \hline  & \\ 
        $x$ & 
        $1.0x10^{-5}$, $1.59x10^{-5}$, $2.51x10^{-5}$, $3.98x10^{-5}$, $6.31x10^{-5}$,\\ &
        $1.0x10^{-4}$, $1.59x10^{-4}$, $2.51x10^{-4}$, $3.98x10^{-4}$, $6.31x10^{-4}$,\\ &
        $1.0x10^{-3}$, $1.59x10^{-3}$, $2.51x10^{-3}$, $3.98x10^{-3}$, $6.31x10^{-3}$,\\ &
        $1.0x10^{-2}$, $1.59x10^{-2}$, $2.51x10^{-2}$, $3.98x10^{-2}$, $6.31x10^{-2}$,\\ &
        $1.0x10^{-1}$, $1.59x10^{-1}$, $2.51x10^{-1}$, $3.98x10^{-1}$, $6.31x10^{-1}$,\\ & $1.0$\\
					     \hline   & \\ 
        $Q^2$ &  $1.0x10^{0}$, $1.78x10^{0}$, $ 3.16x10^{0}$, $5.62x10^{0}$,\\ 
         &  $1.0x10^{1}$, $1.78x10^{1}$, $ 3.16x10^{1}$, $5.62x10^{1}$,\\ 
         &  $1.0x10^{2}$, $1.78x10^{2}$, $ 3.16x10^{2}$, $5.62x10^{2}$,\\ 
         &  $1.0x10^{3}$, $1.0x10^{4}$\\ 
         & \\
          \hline  & \\ 
        $z$ & 0., 0.05, 0.1, 0.15, 0.2, 0.25, 0.3,\\ 
        & 0.4, 0.5, 0.6, 0.7, 0.8, 0.9, 1.0\\
        & \\
        \hline & \\ 
        $P_T$ &  0, 0.05, 0.1, 0.2, 0.3, 0.4, 0.5, \\ 
        &0.6, 0.7, 0.8, 0.9, 1.0, 1.5, 2.0, 4.0\\
        &\\
          \hline
         
    \end{tabular}
    \caption{Kinematic bin boundaries in the main 4-dimensional binning used for the unpolarized TMD evaluations.}
    \label{tab:binning}
\end{table*}

%% file: 3mults.tex
Given that the cross sections vary quite drastically with both $x$ and $Q^2$, it is often useful not to display the actual cross sections for SIDIS measurements but rather multiplicities. In multiplicities the yield of SIDIS events in a particular $x$, $Q^2$, $z$ and $P_T$ bin is normalized by the yield of DIS events in the same $x$ and $Q^2$ bin. Effectively, this normalizes the SIDIS cross sections with their matching DIS cross section and rather highlights the fragmentation aspects of the measurements (although the transverse momentum is of course still a convolution of the participating intrinsic transverse momenta from distribution and fragmentation functions while in the TMD factorization regime). For the full 4-dimensional theoretical extraction, however, cross sections differential in all 4 variables are necessary. 

\begin{figure*}
    \centering
    \includegraphics[width=0.95\textwidth]{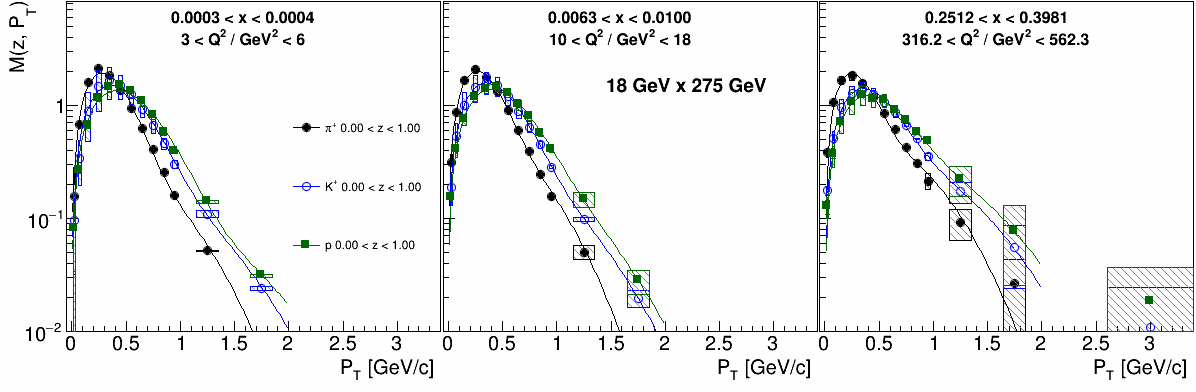}
    \caption{Pion (black), kaon (blue) and proton (green) multiplicities as a function of transverse momentum, in three example bins of $x$ and $Q^2$ with 18 x 275 GeV e+p collisions. For better visibility all $z$ bins were combined. The uncertainty boxes represent the differences between true and smeared multiplicities and serve as a crude estimate of the size of potential systematic uncertainties due to detector smearing and particle mis-identification. The corresponding lines represent fits of double-Gaussians to the multiplicities.}
    \label{fig:multssample}
\end{figure*}

The multiplicities can be seen in Fig.~\ref{fig:multssample} extracted as a function of the transverse momentum for pions, kaons and protons in three example $x$ and $Q^2$ bins to highlight their behavior. The differences between true and reconstructed multiplicities are assigned as a systematic uncertainties which are displayed as uncertainty boxes. This procedure highlights the amount of smearing in the kinematic variables and clearly overestimates the actual uncertainties as smearing and particle mis-identification would in reality be unfolded. However, the regions where large smearing occurs will eventually also show increased uncertainties due to the unfolding while not necessarily being as large as shown in this crude estimation. 
 One can see peaking and then rapidly falling multiplicities that resemble the Gaussian behavior generally seen in fixed target SIDIS measurements at relatively small transverse momenta. In this example figure, therefore a simple double-Gaussian is fitted to the multiplicities (taking only the statistical uncertainties into account) and shows a reasonable description of the pseudo-data. It is interesting to note that pions have the narrowest distribution in all three sample bins while both kaons and protons appear to be wider. Such a feature has been seen also in the Belle data \cite{Belle:2019ywy} and is in part described by {\sc pythia}, but the ordering is different, since in the $e^+e^-$ case (in data and simulation) the differences were most pronounced between mesons and baryons. With the EIC also the transition into the higher transverse momenta, where collinear factorization is needed can be studied in detail.

\begin{figure*}
    \centering
    \includegraphics[width=0.95\textwidth]{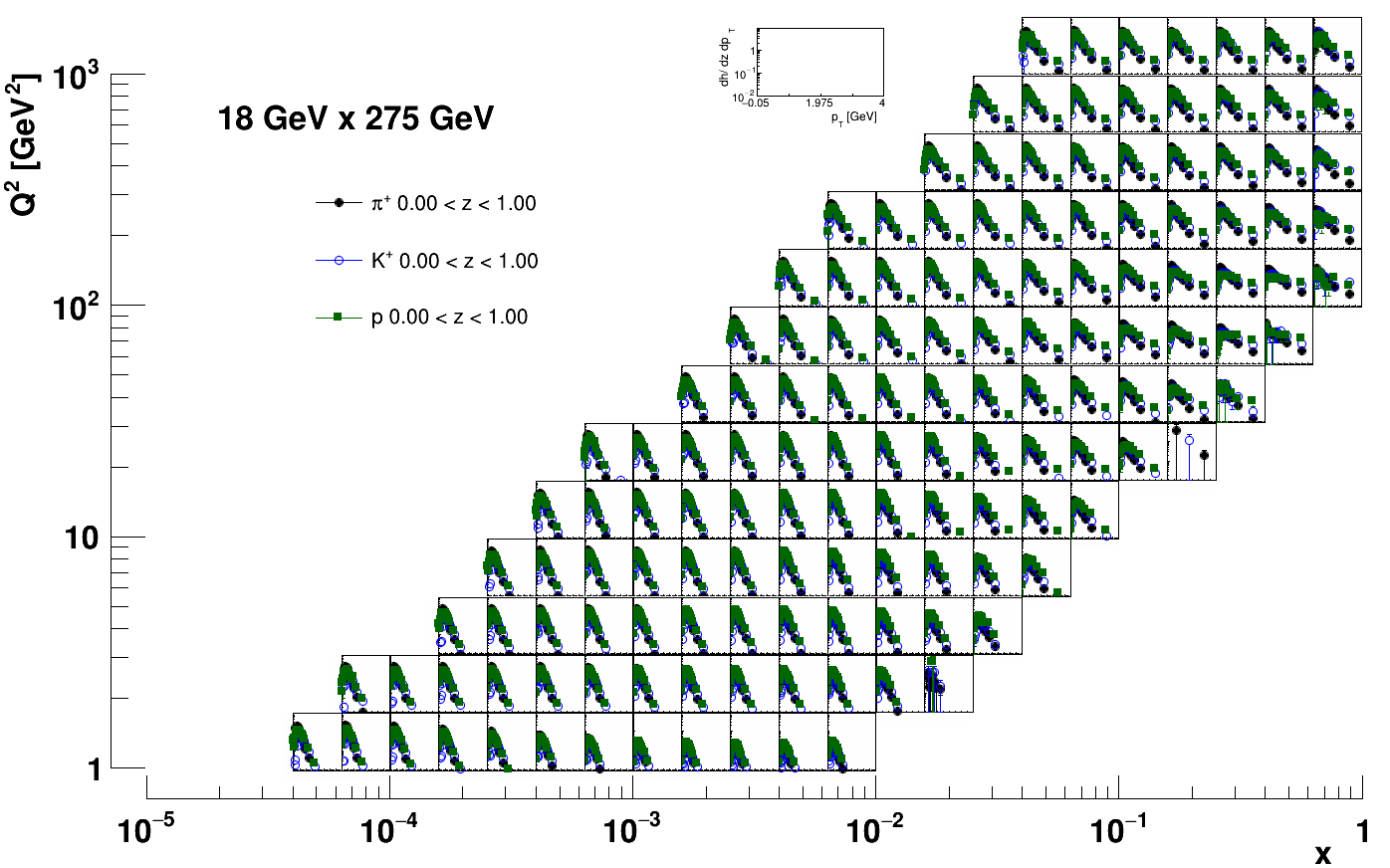}
    \caption{Pion, kaon and proton multiplicities as a function of transverse momentum, in bins of $x$ and $Q^2$ with 18 x 275 GeV  ep collisions. For better visibility all $z$ bins were combined.}
    \label{fig:mults}
\end{figure*}

In Fig.~\ref{fig:mults} the pion, kaon and proton multiplicities are shown for all $x,Q^2$ kinematic bins as a function of $P_T$.
As can be seen, the scale of the multiplicities is very similar for all $x$ and $Q^2$ bins, highlighting the normalization of the rapidly changing cross sections in these variables. A change in the shapes is also visible where generally lower $x$ and $Q^2$ bins show narrower distributions than the at higher $x,Q^2$ bins where the distributions appear to be wider. This figure also highlights the large range that can be covered at the EIC. 

\begin{figure*}
    \centering
    \includegraphics[width=0.95\textwidth]{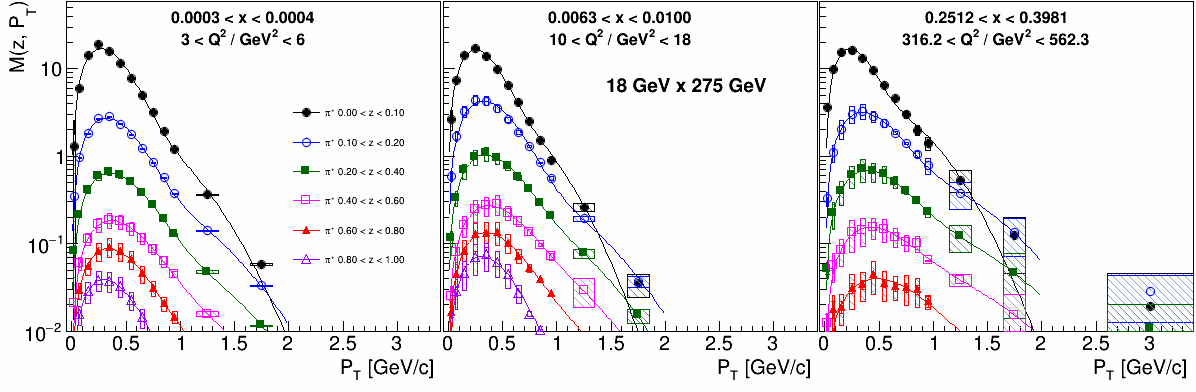}
    \caption{Pion multiplicities as a function of transverse momentum, in bins of $x$ and $Q^2$ with 18 x 275 GeV e+p collisions. Several $z$ bins are shown here to study how the shapes change with $z$ .} The uncertainty boxes represent the differences between true and smeared multiplicities and serve as a crude estimate of the size of potential systematic uncertainties due to detector smearing and particle mis-identification. The corresponding lines represent fits of double-Gaussians to the multiplicities.
    \label{fig:multszsample}
\end{figure*}

\begin{figure*}
    \centering
    \includegraphics[width=0.95\textwidth]{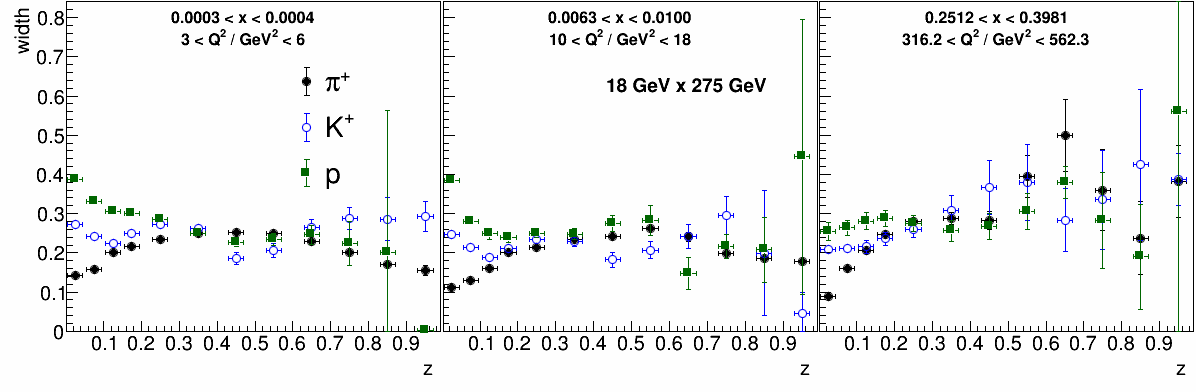}
    \caption{Gaussian widths for pions (black), kaon (blue) and proton (green) as a function of momentum fraction $z$, in example bins of $x$ and $Q^2$ with 18 x 275 GeV e+p collisions. }
    \label{fig:widthzsample}
\end{figure*}

As the multiplicities highlight the fragmentation aspects of the TMD cross sections, it is also very interesting to study the $z$ dependence together with the $P_T$ dependence. This is shown for pions in Fig.~\ref{fig:multszsample} in a few $z$ bins and for three example $x,Q^2$ bins. 
As the transverse momentum that is available in the fragmentation strongly depends on $z$, one should see that the shapes are different in different $z$ bins, appearing narrower at low and high $z$ with a wider distribution at intermediate $z$. In reality this is relatively hard to infer directly from this figure as the scale for the various $z$ bins is vastly different due to the rapidly changing fragmentation functions. As such also double Gaussians are fitted to these sample bins. The widths of the narrower Gaussian, that most likely represents the TMD part of the distribution, indeed shows an increase with $z$ as has been found both in fixed target SIDIS experiments as well as by Belle as is displayed in Fig.~\ref{fig:widthzsample} for all three hadron types. One also sees that the supposedly larger widths from kaons and protons really only appear at low $z$ where especially for protons contributions from target fragmentation may cause these differences.    

\begin{figure*}
    \centering
    \includegraphics[width=0.95\textwidth]{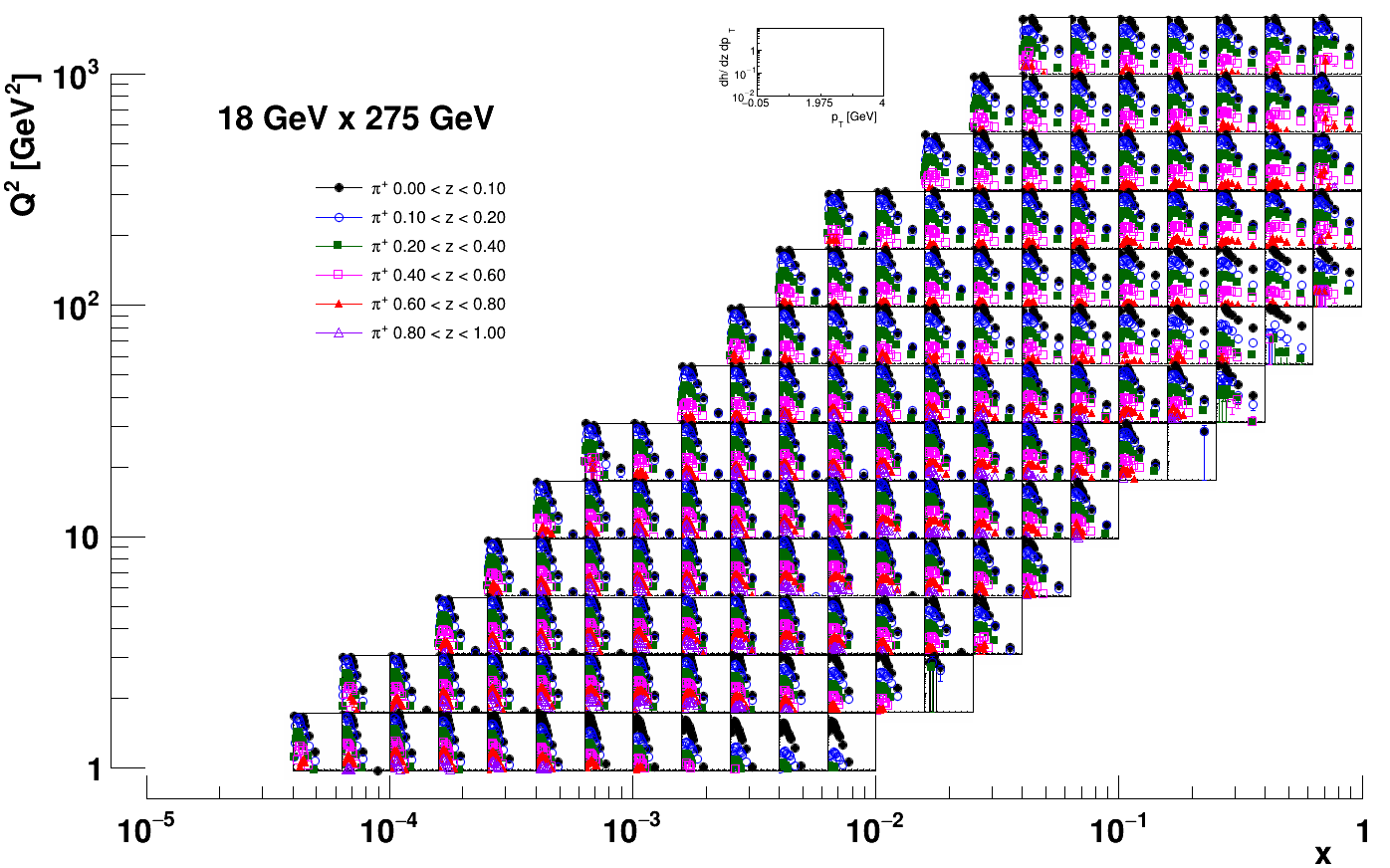}
    \caption{Pion multiplicities as a function of transverse momentum, in bins of $x$ and $Q^2$ with 18 x 275 GeV ep collisions. Several $z$ bins are shown here to study how the shapes change with $z$ .}
    \label{fig:multsz}
\end{figure*}

The whole range for of the covered phase space is shown for pions in various $z$ bins in Fig.~\ref{fig:multsz} where again it can be seen that the multiplicities can extracted in all four kinematic variables with high precision. 

%% file: 4xsecs.tex
\subsection{Cross sections}
\begin{figure*}[th]
    \centering
    \includegraphics[width=0.95\textwidth]{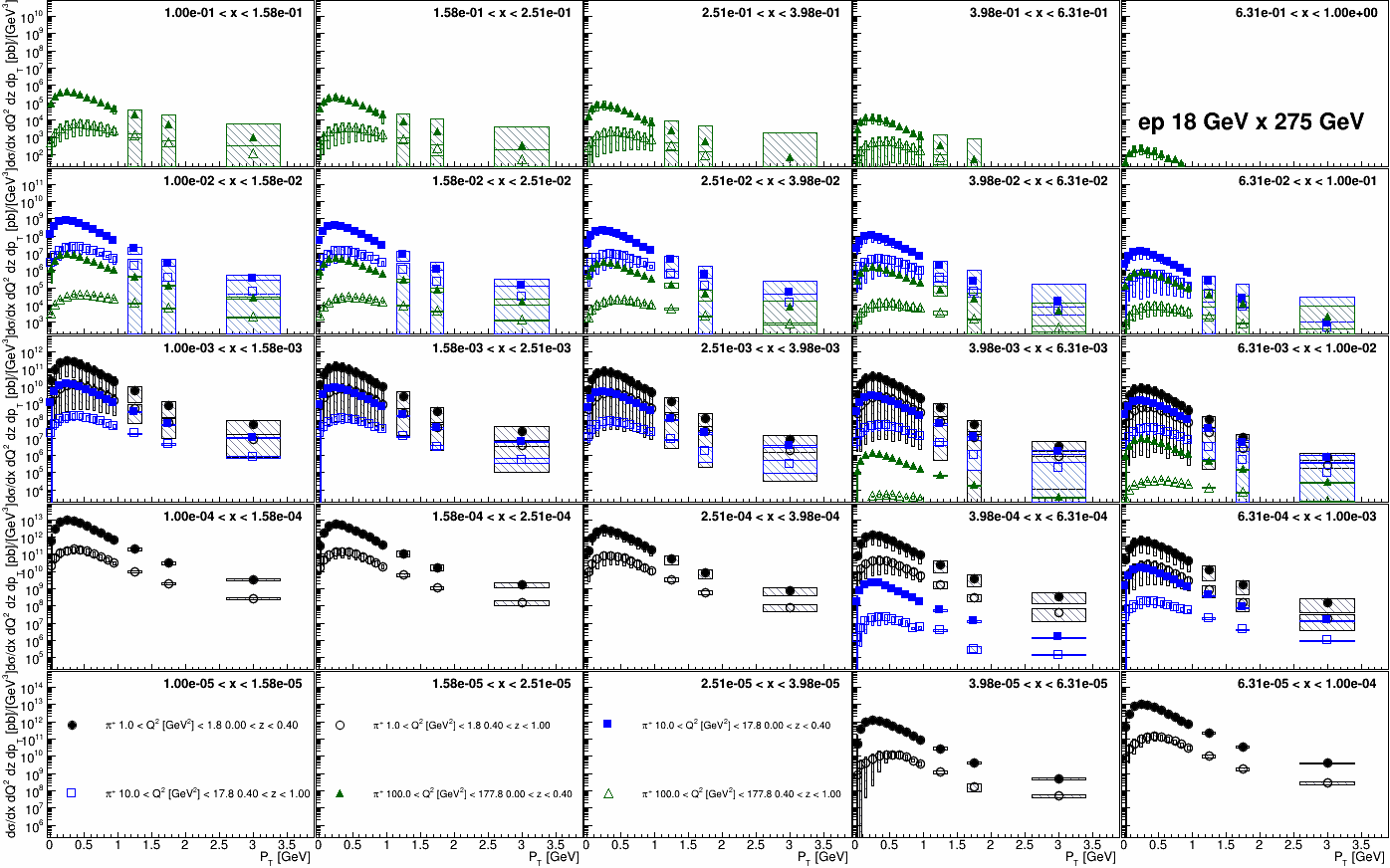}
    \caption{Pion cross sections as a function of $P_T$ in bins of $x$ and for selected bins of $Q^2$. For visibility $z$ bins were combined into the ranges $0<z<0.4$ (full symbols) and $0.4<z<1.0$ (open symbols). The uncertainty boxes are based on the differences between true and reconstructed yields and give an indication of the maximal size of uncertainties due to kinematic resolutions.}
    \label{fig:xsec_pi}
\end{figure*}

\begin{figure*}
    \centering
    \includegraphics[width=0.95\textwidth]{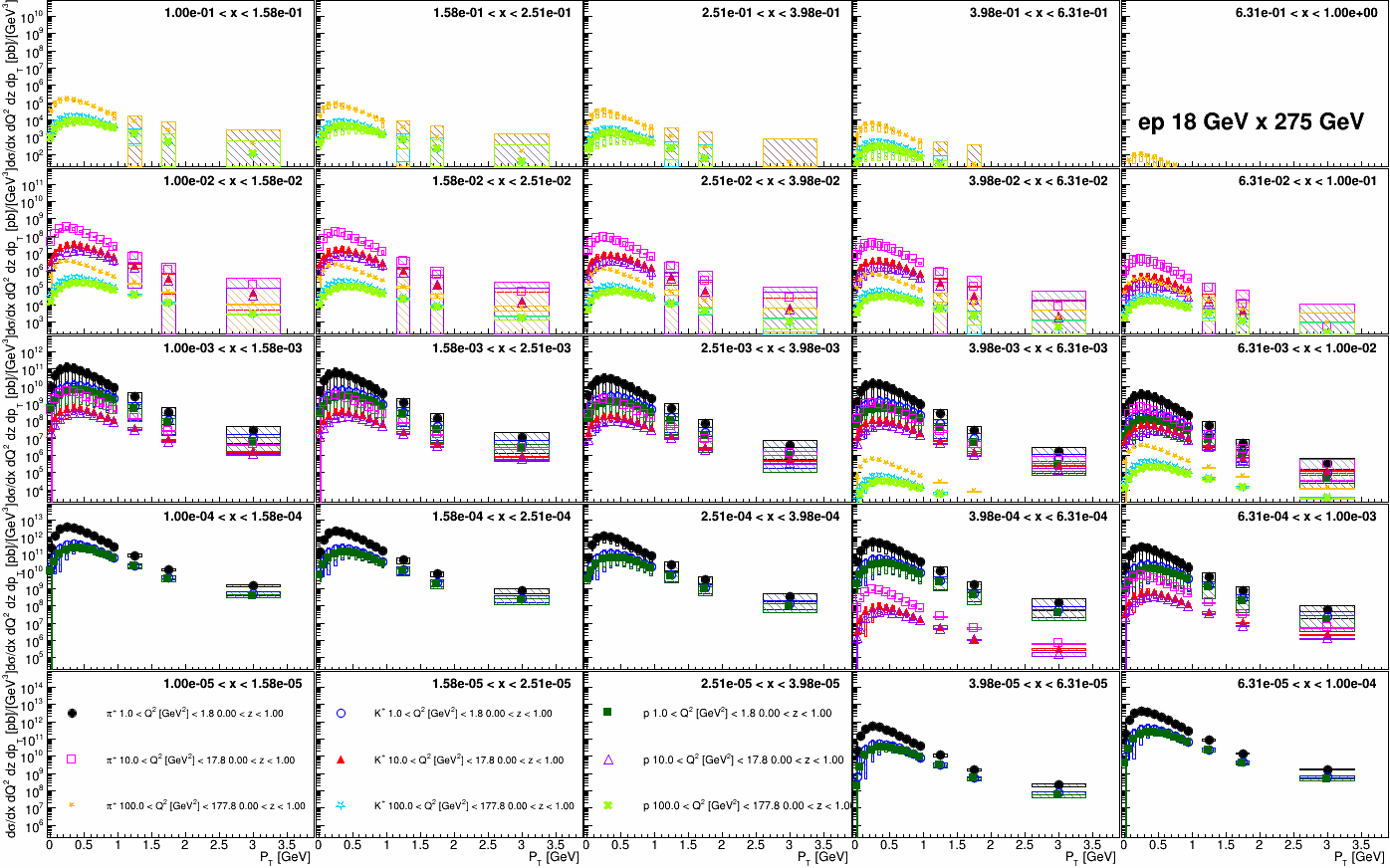}
    \caption{Pion, kaon and proton cross sections as a function of $P_T$ in bins of $x$ and for selected bins of $Q^2$. For visibility  all $z$ bins were combined. The uncertainty boxes are based on the differences between true and reconstructed yields and give an indication of the maximal size of uncertainties due to kinematic resolutions.}
    \label{fig:xsec_pikp}
\end{figure*}

The cross sections for select ranges in $z$ and select $Q^2$ bins are shown in Figs.~\ref{fig:xsec_pi} and \ref{fig:xsec_pikp} for the highest collision energies 18 GeV on 275 GeV. One can see that over a very large range in $x$ the cross sections as a function of $P_T$ can be obtained not only in the range where the transverse momenta are non-perturbative, but also to higher transverse momenta. Already from the simulations one can see some shifts in the shapes of the distributions for low and high fractional energies $z$. 

The systematic uncertainty boxes in these figures represent the differences between true and reconstructed yields and therefore provide a rough measure of how large detector smearing in all the relevant kinematic variables is. It also provides an indication of the maximal size of systematic uncertainties due to these effects. As discussed in the note about SIDIS kinematic resolutions, one also notices here that these uncertainties tend to be larger at the higher $x$ end of the phase space for a given $Q^2$ bin which corresponds to the lower $y$ region where the DIS kinematic reconstruction via the scattered lepton is less precise. Also at higher $z$ and $P_T$ this behavior sets in earlier than at lower values of these variables which relates to generally slightly larger smearing at higher values. 

\begin{figure*}
    \centering
    \includegraphics[width=0.95\textwidth]{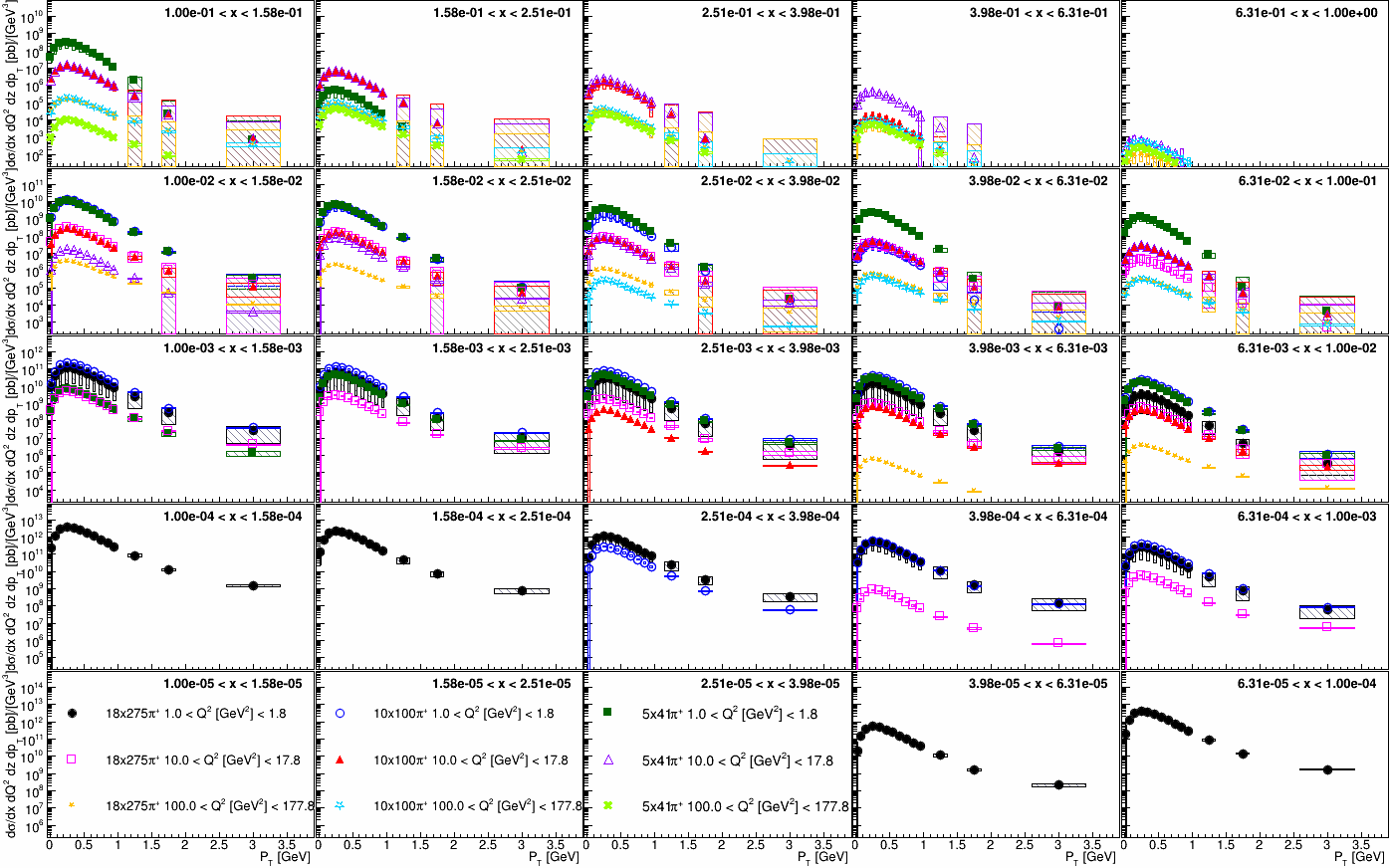}
    \caption{Pion cross sections as a function of $P_T$ in bins of $x$ and for selected bins of $Q^2$ for three different collision energies. For visibility all $z$ bins were combined. The uncertainty boxes are based on the differences between true and reconstructed yields and give an indication of the maximal size of uncertainties due to kinematic resolutions.}
    \label{fig:pienergy}
\end{figure*}

From these example figures one can see that already within one beam energy one can reach a reasonable range in $Q^2$ for the same $x$ bin which will significantly improve the understanding of the TMD evolution. Note that these cross sections are coming from a MC simulation and the actual magnitude may be rather different. Nevertheless, even in the very limited luminosities that were simulated, the statistical precision up to high $x$ and $Q^2$ is sufficient that even in the case of decreased cross sections due to evolution or radiative effects these questions can be addressed. 
Additionally, using different collision energies, one can further augment this sensitivity as the lower collision energies allow to reach higher $x$ at lower scales than at the high collision energies. This is illustrated in Fig.~\ref{fig:pienergy} where select $Q^2$ bins are shown for pions from different collision energies. One can see that at intermediate $x$ and $Q^2$ all overlap while at the more extreme regions either the highest or the lowest collision energies still have coverage.

\subsection{Projections}
Similar to the Yellow report, the expected cross sections in the described four-dimensional binning in $x$, $Q^2$, $z$ and $P_T$, with statistical uncertainties scaled to 10 fb$^{-1}$ for each collision energy were then provided to the theorists for the impact studies on the unpolarized TMD distribution and fragmentation functions, as well as the non-perturbative parts of the TMD evolution. The systematic uncertainties are again estimated by comparing the true and reconstructed yields and therefore can be considered an upper limit on the systematics due to detector effects. They are generally expected to be the dominating uncertainties even after unfolding and likely smaller than the uncertainties due to luminosity and acceptance evaluation. 

The theortical groups then use the pseudo-data to evaluate the impact that data can have on the uncertainties of their fits. They generally cannot perform actual new fits of the pseudo-data since the precision and number of data points far exceeds the existing data and the length actual new fits would take. Instead, they generally re-evaluate their existing fits by re-weighting their sets of samples or replicas including the uncertainties of pseudo-data, cf. \cite{Giele:1998gw,Ball:2010gb} and a discussion of these methods in \cite{Sato:2013ika}. It should be noted that this means that they can only estimate the improvement in precision within their parametrizations.   
From the Yellow report, the Pavia group \cite{Bacchetta:2019sam}, for example, estimated the relevance of the EIC unpolarized TMD data on the different parts of the TMDs. 
They found that the nonperturbative contributions to the TMD evolution get addressed over a wide range of the phase space, highlighting the required lever arm in $Q^2$ to study evolution. Naturally, the low-$x$ parts of TMDs were so far hardly accessible, so their knowledge also gets substantially improved by the EIC. Also the fragmentation related quantities receive significant improvements. 
One aspect that cannot be addressed using the re-weighting is how the increase in data-points and the related increase in sensitivity will affect aspects that are not covered in the current parametrizations. Given the limitations of the exisiting data, so far only are very limited dependence on flavors is included in the the fits while the expectation is that the EIC data can actually determine at least light quark valence and sea flavors in the TMD PDFs and FFs. 


%% file: 5impact.tex
\begin{figure*}
    \centering
    \includegraphics[width=0.9\textwidth]{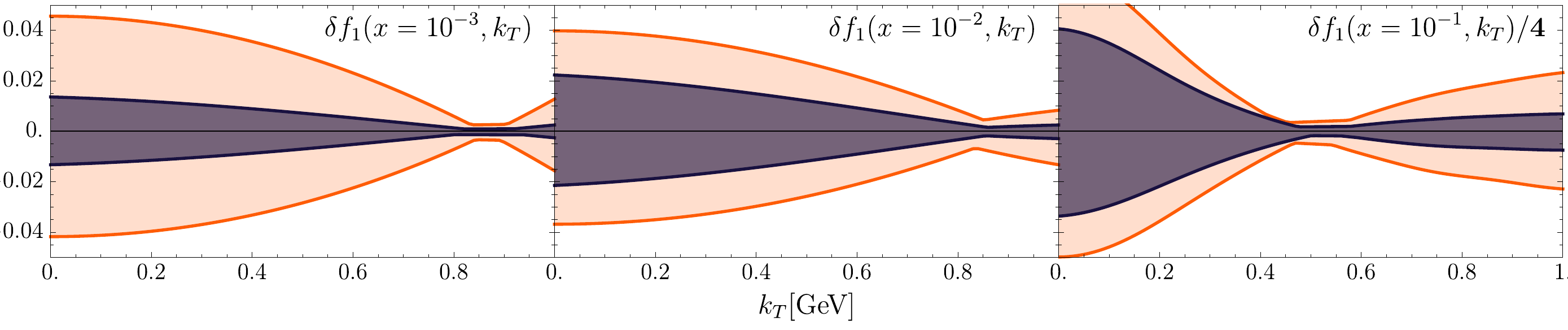}
    \includegraphics[width=0.9\textwidth]{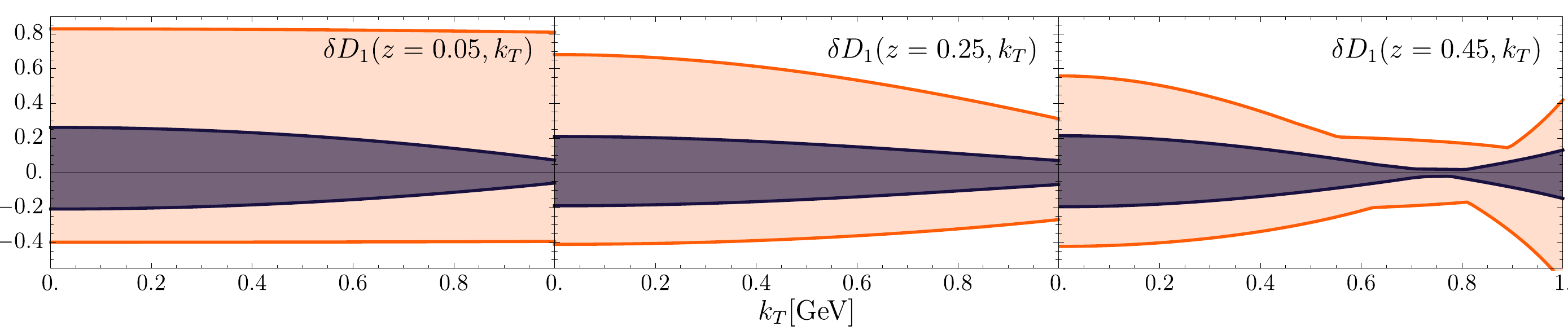}
    \caption{Expected EIC uncertainties on the unpolarized TMD PDFs (top) and FFs (bottom) as a function of the intrinsic transverse momentum for certain $x$ and $z$ slices in comparison to the existing uncertainties.}
    \label{fig:unpolTMDs}
\end{figure*}

Once the pseudo-data is analyzed one can obtain the relevant improvements on the TMD PDFs and FFs within the framework of the existing global fits, as discussed in the previous section. As an example the current and expected uncertainties following the unpolarized extraction of Ref.~\cite{Scimemi:2019cmh} are shown in Fig.~\ref{fig:unpolTMDs}.
These show for selected $x$ and $z$ slices the expected impact on the intrinsic transverse momentum dependence of the distribution and fragmentation functions when including the EIC pseudo-data. One can see that the impact is quite substantial. It also shows some of the limitations of the re-weighting approach as in the distribution functions a node-like feature is visible that originates from the functional form and the overall normalization of this particular global fit. 
Nevertheless, it highlights the impact the EIC data will have on the TMDs and their theoretical understanding, including the scale evolution of TMDs, eventually their flavor dependence and finally the three-dimensional momentum picture of the nucleon for various parton flavors.

%% file: 6outlook.tex
In conclusion, the unpolarized transverse momentum dependent hadron multiplicities as well as cross sections can be extracted over a large range in the DIS kinematic variables $x$ and $Q^2$ and the semi-inclusive variables $z$ and $P_T$. The ECCE detector configuration is well suited to obtain a comparable precision as the reference detector parametrization used in the studies of the Yellow Report. 
From these measurements the precision on the transverse momentum dependent distribution and fragmentation functions can be greatly increased, likely allowing for a detailed, flavor dependent extraction of these TMDs. The lever arm of these measurements will definitely remove the uncertainties that currently exist on the TMD evolution. Also, further theoretical insights into the regions of applicability of TMD factorization, collinear factorization, etc can be explored in a large area of phase space.  